\documentstyle[11pt]{article}
\def\beq{\begin{eqnarray}}
\def\eeq{\end{eqnarray}}
\def\bea{\begin{eqnarray*}}
\def\eea{\end{eqnarray*}}

\def\centeron#1#2{{\setbox0=\hbox{#1}\setbox1=\hbox{#2}\ifdim
\wd1>\wd0\kern.5\wd1\kern-.5\wd0\fi
\copy0\kern-.5\wd0\kern-.5\wd1\copy1\ifdim\wd0>\wd1
\kern.5\wd0\kern-.5\wd1\fi}}
\def\ltap{\;\centeron{\raise.35ex\hbox{$<$}}{\lower.65ex\hbox{$\sim$}}\;}
\def\gtap{\;\centeron{\raise.35ex\hbox{$>$}}{\lower.65ex\hbox{$\sim$}}\;}

\def\singleandthirdspaced{\baselineskip=\normalbaselineskip\multiply
    \baselineskip by 130\divide\baselineskip by 100}

\newcommand{\newc}{\newcommand}
\newc{\qbar}{{\overline q}}
\newc{\Kahler}{K\"ahler }
\newc{\deltaGS}{\delta_{\rm GS}}
\begin{document}
\begin{titlepage}
\begin{flushright}
{\large 
 SCIPP-2008/06\\
}
\end{flushright}

\vskip 1.2cm

\begin{center}

{\LARGE\bf Symmetric Points in the Landscape as Cosmological Attractors}

\vskip 1.4cm

{\large Michael Dine, Guido Festuccia and Alexander Morisse}
\\
\vskip 0.4cm
{\it Santa Cruz Institute for Particle Physics and
\\ Department of Physics, University of California,
     Santa Cruz CA 95064  } \\
\vskip 4pt

\vskip 1.5cm

\begin{abstract}
In the landscape, if there is to be any prospect of scientific prediction, it is crucial
that there be states which are distinguished in some way.  The obvious candidates are
states which exhibit symmetries.  Here we focus on states which exhibit
discrete symmetries. Such states are rare, but one can speculate
that they are cosmological attractors.  We investigate the problem
in model landscapes and cosmologies which capture some of the features of
candidate flux landscapes.  In non-supersymmetric theories
we find no evidence that such states might be cosmologically
favored.  In supersymmetric theories, simple arguments suggest
that states which exhibit $R$ symmetries might
be.  Our considerations lead us to
raise questions about some popular models of eternal inflation.
\end{abstract}

\end{center}

\vskip 1.0 cm

\end{titlepage}
\setcounter{footnote}{0} \setcounter{page}{2}
\setcounter{section}{0} \setcounter{subsection}{0}
\setcounter{subsubsection}{0}

\singleandthirdspaced

\section{Introduction:  Distinguished States in the Landscape}

For almost 30 years, inspired by string theory, particle theorists have imagined
that there is a sensible program of attempting to infer the laws of nature from
some grand set of principles -- from the top down.  The emergence of the string
landscape, however, has thrown this paradigm into question.
If there is any truth to the idea of a cosmic landscape\cite{weinberglambda,bp,
kklt,lennylandscape,
lennybook}, there is no realistic
program in which one finds {\it the} string ground state which describes
the world around us.  At best, one can hope to do statistics, looking for strong
correlations between known facts and laws of nature which can be studied
in future experiments.

At present, even an enthusiast must acknowledge that the landscape is more of
a vision than a well-defined theoretical structure.  At best, we have explored
only a small subset of the possible set of states\cite{IIB,IIA}; most of these are
supersymmetric or approximately so.   The most interesting
states, with broken supersymmetry and small cosmological constant, can hardly be said to have
been understood in a systematic way.  In the original KKLT
construction, for example, while there are small numbers, there is not
a formal, small parameter in which one can develop a systematic
approximation scheme.  On the other hand,
the KKLT analysis seems plausible, and the
{\it statistics} which arise from this and similar constructions seem robust.  For example, the statistics discussed
by Douglas and Denef\cite{douglasdenefsusy,douglasdenefnonsusy}
emerge from simple assumptions about the
low energy degrees of freedom, as well as assumptions of uniform distributions
of lagrangian parameters with respect to some
natural measure\cite{dos}.
So it is perhaps not unreasonable to think that these {\it model landscapes}
may provide reliable insights into the larger structure.

It is not
at all clear whether the (nearly) supersymmetric states are representative; serious arguments
have been advanced that they are not\cite{douglasdenefnonsusy,silverstein}.
So, even accepting the validity of the landscape framework, it is unlikely
that we will soon be able to reliably enumerate all possible classes of landscape
vacua.  Thus if there is to be any hope of prediction, it must emerge from
some more general considerations.  Among classes of states which have been studied,
those which might be distinguished are:
\begin{enumerate}
\item  (approximately) supersymmetric states
\item  states with large internal volume
\item  states exhibiting warping
\item  states exhibiting (discrete) symmetries
\end{enumerate}

While most thinking about states of string theory assumes some approximate supersymmetry,
it seems reasonable to think that supersymmetry is special, and that there might be
vastly more supersymmetric states than non-supersymmetric states\cite{silverstein,douglasdenefnonsusy,douglassusy,
lennynonsusy}.
One can ask, on the other hand, whether cosmologically, such states might be favored.
In \cite{dfmv}, it was pointed
out that primitive cosmological considerations might favor approximately
supersymmetric states.  Typical stationary points
of an effective action are extremely unstable -- they
are not states at all.  Approximately
supersymmetric states,
however, are inevitably highly metastable.  Of the various distinguishing
characteristics listed above, apart from supersymmetry, only large volume generically provides
stability, and, at least within IIB landscapes, large volume does not appear
generic.  We won't say more about large volume or warping in this paper, as our
considerations don't seem to provide any arguments that these states are special.

In speculating about physics beyond the Standard Model, it has
been a widely held belief that discrete
symmetries might plausibly -- and naturally --
solve a range of problems.  Such symmetries often arise
at special points in the moduli spaces of critical string constructions.
But in the framework of the flux landscape, discrete symmetries would seem to be rare\cite{dinesun}.
The problem is that the vast number of states purported
to exist within the landscape arise
because of the existence of a large number of fluxes which can take a large number
of values.  However, in order to obtain symmetries, many fluxes must be set
to zero, greatly reducing the ``dimensionality" of the space.  In other words,
only an exponentially small fraction of states exhibit symmetries.

Again, we can ask whether cosmology might in some way favor (or not) states
exhibiting discrete symmetries.  Here, the issue is not one of
stability, but rather:  are the symmetric states, within some
realization of eternal inflation, often attractors.  We might mean
by this that the measure in some inflationary model is highly peaked
at the symmetric states, but we will shortly
formulate the question in a more naive and primitive way.

Given that there is so much we do not understand about the landscape, the burden of this  paper is not to {\it prove}
that symmetric states are attractors, but rather to establish that this possibility is, or is not, plausible.
For this, we seek to adopt models of the states within the landscape, the transitions between them, and
the mechanisms by which they are populated.   For the states, the challenge
is to study the highly non-supersymmetric states which are the antecedents of the would-be states of interest.
We will consider a number of models for the states,
starting with BP, and then continuing to ``states" found as stationary points of supergravity models.  These models
will incorporate some, but not all, of the features we might expect to hold in ``realistic" theories.
In considering transitions, as we have indicated, we will suppose radii and couplings to be such that these
states {\it are} states, but that they are highly metastable, so that they do not experience periods of exponential
growth.   One of our more drastic assumptions will be to treat the fluxes as quasi-continuous.  In this way,
we can think of tunneling as giving rise to a flow within the landscape, and ask whether, for a given starting
point, the system flows towards symmetric states.  Such a continuous flux approximation has proven useful
in thinking about landscape statistics\cite{ashokdouglas}, and has been applied in other cosmological contexts\cite{shenker}.
While surely unrealistic for this problem, it is hard to believe that if symmetries are {\it not} attractors in this framework
that they are in any realistic setting.

Finally, we need a model for cosmology.  Ultimately, we must face up to questions
of measures for eternal inflation.  But our goal, here, is simply to establish whether
there is any reason to think that the rare states in a landscape exhibiting
symmetries are somehow favored.  For this, again, the burden of this paper
is not to establish conclusively that this is the case in an underlying, complete
theory of gravity, but simply to establish some general conditions under which
symmetries might plausibly be favored.
Postulating a landscape with a large number of (very) metastable de
Sitter states,  we will consider possible ``initial conditions"
in which the universe starts in one such state. Then we can ask whether, for a non-negligible subset of
possible starting points, the system finds its way to the symmetric
state.  To model this problem, we will suppose that the antecedents of
the symmetric state are short lived and
do not experience long periods of exponential growth.

In the next section, we lay out certain basic issues involving symmetries in such
a picture.    We introduce the notion of a neighborhood in the landscape, and argue
that it is plausible that an order one fraction of states lie in the neighborhood of states
exhibiting discrete symmetries.  In section three, we present simple arguments
for our basic results:  ordinary symmetries are not attractors, but R symmetries
might be.   We survey a set of model landscapes in section
four.  In this discussion, we enumerate some of the present obstacles
to developing a landscape cosmology.  We discuss the problem
of symmetric points within model landscapes.  The $T_6/Z_2$ orientifold
is introduced as a model where the symmetry structure is readily understood.
Our studies reinforce the arguments of section two, that points exhibiting
$R$ symmetries can readily be attractors, but other types of symmetries
are not singled out.

In section five, we discuss our conclusions.
We explain
that these studies are, at best, preliminary and suggestive.  Making definitive statements
about cosmology certainly requires a deeper understanding of non-supersymmetric
configurations within the landscape.  But our discussion, we believe, makes clear
that the viewpoint, advanced by some anthropic advocates (e.g. \cite{halletal}),
that symmetries play no role, does not necessarily follow  from adopting a landscape framework.
There are clear ways in which supersymmetry and discrete symmetries might well
emerge from cosmological considerations.  Obviously continuous gauge symmetries
are of great interest; we leave their study for future work.

\section{Neighborhoods of Symmetric Points}

States exhibiting discrete symmetries are rare in the flux
landscape.  But the fraction of states lying nearby
such symmetric points need not be small.
To understand this, suppose we have some number of symmetric fluxes, $N_i$,
$i=1,\dots I$, and symmetry-violating fluxes, $n_a$, $a = 1 \dots A$.  We want
to ask whether the set of states in the neighborhood of the symmetric points,
i.e., in which $n < N$, is an order one fraction of the total
number of states -- or more precisely, initial conditions.  To model this, let's
suppose that the initial state is taken to lie on the ``hypersphere"
\beq
N_i^2 + n_a^2 = R^2
\eeq
(with summation on $i, a$) and ask what fraction of the volume of the sphere satisfies
\beq
\alpha^2 N_i^2 > n_a^2.
\eeq

For simplicity, we'll take the measure in the space of fluxes to be flat.
So we need to compare:
\beq
\int d^I N d^A n \delta(\sqrt{N^2 + n^2} - R^2)
\eeq
with the same integral, restricted by $n^2 <\alpha^2 N^2$ (the constant $\alpha$, should
be less than one, but our working assumption
is that it need not be much less than one).
To do the integrals, we need to recall that the solid angle in $d$ dimensions is:
$
\Omega_d = {2 \pi^{d \over 2} \over \Gamma{(d/2)}}.
$
So the integral, in the unrestricted case, is simply
\beq
2 R^{I+A-1} \sqrt{\pi}^{I+A} \over \Gamma({I + A \over 2}).
\eeq
With the restriction on the smallness of $n$, the number of states is
\beq
\int d^I N \int d^A n \delta(\sqrt{N^2 + n^2} - R^2) \theta(\alpha^2\vert N^2 \vert -
\vert n^2 \vert).
\eeq
$$~~~~= {2 R^{I+A-1} \sqrt{\pi}^{I+A}\over \Gamma(A/2)\Gamma(I/2)} \int_0^{\alpha \over
\sqrt{1 + \alpha^2}} dx x^{A-1}(1-x^2)^{{I\over 2}-1}.$$

In the case that $\alpha \ll 1$
we can readily estimate the integral, approximating the $(1-x)^{I/2-1}$ factor in the integrand by
$1$.  Then the integral behaves as $\alpha^{A}$, while the
total number of states behaves roughly as $\left({A\over 2}\right)^{-{I\over 2}}$, so the overall suppression is
$\alpha^{A} \left({A\over 2}\right)^{I\over 2}$.
For large $A$ and small $\alpha$, this can be very substantial.  If $\alpha \simeq 1$, on the other
hand, numerical study of the
integral indicates that
the suppression with increasing
$A$ is quite modest.  At the moment, the correct value of $\alpha$ is something
we can only guess, but we will assume, in what follows, that it is not
too small.

One interesting question is how much disfavored are larger symmetries.\footnote{We thank
L. Susskind for stressing the significance of this issue.}  As an example,
we consider the case, numerically, where we take $\alpha=1$, and compare $I=2,A=10$, with
$I=6,A=6$.  The difference in the two cases is a factor of $16$, i.e. not huge.  So it is not
clear that there is an overwhelming preference for smaller symmetries (or at least for small
vs. slightly larger symmetries).

\section{Symmetric Points: Attractors Within a Landscape?}
\label{attractors}

Having established that an order one fraction of states might lie in the neighborhood
of a symmetric point, we ask:  with our assumptions about initial conditions, is the system
more likely to end up in a symmetric state than a non-symmetric state?

Consider, first, the general question of tunneling within a flux landscape.
In \cite{dfmv} this question was discussed from a slightly different point of view,
which we will adopt here.  The general idea is that the large number of states in the
flux landscape arise because there are a large number of possible fluxes, $N_i = 1,\dots b$,
with each of the $N_i$ taking on a large number of possible values.  The potential, $V(\phi)$,
is typically quadratic in fluxes (e.g. in the IIB landscape), $V \sim N^2$, so:
\begin{enumerate}
\item  The difference in energy between states differing by a single unit of flux
is of order $N$
\item  The tension of the branes separating states differing by a single unit of flux does not
depend significantly on the total flux.   This follows from
considering wrapped branes (NS5, D5).  Alternatively, from the point of view of the fields
in the low energy theory, the tension
is of order $\sqrt{V(\phi)} \Delta \phi \sim 1$; here $\Delta \phi$, the change in typical
fields (e.g. complex structure moduli in the IIB case) is of order $1/N$.
\item  The bounce action for transitions between neighboring states is typically
small for large $N$, scaling as
\beq
S_b(\Delta N = 1) \sim {1 \over N^3}.
\eeq
\end{enumerate}

Ref. \cite{dfmv} focussed on the question:  if a state happens to have small cosmological
constant, is it stable against decay to its numerous neighbors with negative cosmological
constant.  Here we are asking a different sort of question.  We are studying the
ancestors of the symmetric states.  Since we are interested in the
symmetric states as candidate
theories of nature, we will assume that they have small cosmological constant; the ancestors
have positive cosmological constant.
The estimate above indicates that these states are typically highly unstable -- they
are not states at all.  But, as a model, we will take this type
of formula literally (one can
suppose, for example, that the states, for some reason, all have moderately large volume,
enhancing the bounce action).  In the same spirit, we will treat the fluxes as
large enough that they can be treated as quasi-continuous\cite{ashokdouglas}.
In such a picture, there is a well-defined notion of {\it neighborhood} in the landscape.

As a further, rather drastic simplification, we will ignore gravity in estimating decay amplitudes.
This is often valid (as discussed, for example, in \cite{bp}), for some range of fluxes and values of moduli,
but it is not likely to hold generally.  For example, among the many numerous metastable dS states,
it is likely that some are long-lived, giving rise to (eternal) inflation.
As we will see, even with such drastic assumptions, only in a limited set of instances do symmetric
states seem singled out by cosmological considerations.

\subsection{Non-Symmetric States}

Suppose that the putative state of interest, with small cosmological constant,
lies at some point in the flux lattice $\vec N_0$.   In this case, in the
continuous flux approximation, we would expect the energy, near that point,
to have the form:
\beq
E_0 = C_a \Delta N^a
\eeq
Recall the basic assumption that there is a large, multi-dimensional space of states.
So, from the tunneling perspective,
with our assumption about initial conditions, it is clear that from most starting points,
one will ``miss" the state $\vec N_0$, ending up eventually tunneling to negative cosmological
constant big crunches.  The chance of ending up in the ``correct vacuum" will be of order one
divided by the number of low cosmological constant states.

\subsection{The neighborhood of symmetric states}

Consider a state which is symmetric or approximately symmetric under an
ordinary discrete symmetry.  Some subset of fluxes, $N_i$, $i=1,\dots,B$, respect the
symmetry (they are neutral under the symmetry, and there are minima of the resulting
potential in which only fields neutral under the symmetry have expectation values).
The rest, $n_a$, $a= 1,\dots,A$, break the symmetry.   The putative,
low cosmological constant, symmetric state we denote, again,
by $\vec N_0$.  The set of states with flux
\beq
\vert n_a \vert < \vert N_i -N_i^0 \vert
\eeq
define the neighborhood of the symmetric point.
We will take $n$ to denote an average $n_a$, $N$ an average $N_i$.

If we assume that in the symmetric state there are no moduli, then
in the continuous flux approximation, the energy landscape nearby has the form
\beq
E_0 = \sum_{i,J=1}^I f_{IJ} N_I N_J + \sum_{a,b=1}^A g_{ab} n_a n_b
\eeq
Again, in the continuous flux approximation, we can diagonalize $g$.
We can immediately distinguish two cases.
\begin{enumerate}
\item  $g_{ab}$ has only positive eigenvalues.  In this case, if one starts in the neighborhood
of the symmetric state, at least the transitions which change the $n_a$'s will tend towards
the symmetric state.  A priori, we might simply assign a probability $1/2$ for the possibility that a particular
eigenvalue is negative, so there is (compared to factors like $10^{500}$) a modest
suppression of this possibility.
\item  $g_{ab}$ has some negative eigenvalues.  In this case, the corresponding
$n$'s will tend to grow, and the system will not tend towards the symmetric state.
\end{enumerate}

But even in the case where all of the eigenvalues of $g$ are positive,
there is nothing particularly special about transitions to the
symmetric $\vec n_a =0$ state.  Assuming the eigenvalues of $g$ are all
similar, if $n_a \ll N_i$, then
\beq
{S_b(\delta n_a=1)\over S_b(\delta N_i=1)} \sim {N^3 \over n^3}
\eeq
So transitions which change $n$ are much slower than those which change $N$.
Under such circumstances, the notion of a neighborhood is not particularly
relevant to the tunneling process.  Unless one starts in the state with $n_a=0$,
one will not reach the symmetric states; one may reach a state $(\vec N_0,\vec n_a)$,
but then one will transition to big crunches with negative cc.

Suppose, on the other hand, that the elements of $g$, in addition to being positive,
were far larger that those of $f$.  Then transitions involving changes in
$n$ would be faster than those involving changes in $N$, and states in the neighborhood
would, preferentially, end up in the symmetric state.  Such a situation could
conceivably arise if there were approximate moduli in the symmetric state.  Then,
in the nearby states, some quantities (volumes of cycles, for example) might be
quite different than in the symmetric state, and tunneling rates could be quite
different.  We will a phenomenon of this sort in the case of R symmetric vacua.

\subsection{R Symmetries}
\label{rsymmetries}

Let's turn our attention to $R$ symmetric states.
Here there is an important difference with the non-R case.  Changes in $N$, in the
symmetric limit, are not associated with changes in energy, so they are potentially
highly suppressed.

To go further, we need to think more carefully about the R symmetric states.
In general, suppose that we have fields, $X_I$, which transform
under the $R$ symmetry like the superpotential, and fields, $\phi_\alpha$,
which are neutral.  There are additional fields, $\rho_i$,
which transform differently than the $X$'s under
the $R$ symmetry.  Then the superpotential has the form
\beq
W = X_I F_I(\phi_\alpha,\rho_i)
\eeq
There will be supersymmetric, $R$ symmetric solutions ($X=\rho=0$) provided there
are more $\phi$ type fields than $X$ type fields.
In this case, in general, there will be some number of moduli.
This condition holds in many Calabi-Yau compactifications, for example\cite{dinesun}.

In the present context, this means that classically the vacua
with $n_a=0$ possess moduli.  Our working assumption
will be that these moduli are fixed by non-perturbative
dynamics, as discussed in \cite{branches}, with
supersymmetry broken at comparatively low scales.
For non-zero $n_a$, the situation is quite different.
We can study an effective action
for the moduli.  Their potential is of order $n^2$.  Correspondingly,
the masses-squared of the light moduli are of order $n^2$.  As we will
see in model examples shortly, the corresponding ground states can be
dS or AdS.  If much of the neighborhood of the symmetric point is AdS,
we don't expect a significant probability to reach the symmetric
point.  So we will assume that the neighbors of the symmetric point
are predominantly dS.

In order to compare tunneling rates for processes with changes in $N$ with those
with changes in $n$, we need to understand how the properties of the lagrangian
for the light moduli
depend on $N$.  We expect that this dependence is quite weak; indeed,
we might guess that it goes as
$n^4/N^2$.  We can illustrate the point, again, with a simple field theory model.
Consider a theory with three fields, $X$ and $\phi, \chi$, where $X$ transforms like $W$ and
$\phi,\chi$, are neutral.  For the superpotential we take
\beq
W = \lambda X (\phi + \chi)  (\phi-\chi)
\eeq
(the precise dependence on $\phi$ and $\chi$ is not crucial, but makes the equations
simple).  There is a moduli space, on one branch of which
\beq
\phi = \chi = v
\eeq
On this branch, there are two massive fields, $X$ and $\Phi= {1 \over \sqrt{2}}
(\phi - \chi)$, and one massless field,
$\psi = {1 \over \sqrt{2}} (\phi + \chi)$.
Now, the effect of some underlying theory in which we turn on fluxes, $n_i$, would be
to generate a superpotential of the form, say
\beq
\Delta W = \epsilon (\alpha + \beta \psi + \gamma \Phi).
\eeq
where $\epsilon \sim n$.
Such a potential can fix the remaining modulus (and
break supersymmetry).  It also generates a tadpole for $\Phi$,
of order $\epsilon^2$, and a $\lambda$-dependent term in the energy of order
\beq
\Delta E \approx {\epsilon^4 \over \lambda^2 v^2} \sim {n^4 \over N^2}.
\eeq
We will see some of these features in more detailed models in subsequent sections.

We will be interested in the relative size of transition rates for changes
in $n$ and changes in $N$.  We have understood the energy splittings
between the different types of states; we need to ask about the tensions.  One might
think that the tensions for changes in $N$ would also be suppressed, as, for BPS configurations,
the tension is proportional to $\Delta W$.  But already in field theory, domain walls separating
$R$ symmetric states are not BPS.  As a simple example, consider
a theory with superpotential
$$W= X(\phi^2 - \mu^2)$$
This model has an $R$ symmetry and a $Z_2$ symmetry.  The states $\phi = \pm \mu$ are separated by a region in which
$\phi \ne 0$, and ${\partial W \over \partial X} \ne 0$, but in which $\partial_\mu X=0$, so they are not BPS,
and their tension is certainly non-vanishing.  In the flux case, we expect, again, that the tension is not very sensitive to $N$.
So the bounce action for changes in $N$
is larger than that for changes in $n$ by
\beq
{S_b(\Delta N =1) \over S_b(\Delta n =1)} = {N^9 \over n^9}!
\eeq
So in this case, transitions towards the symmetric point are likely to be much
faster than other transitions.

This discussion, while suggestive, is not complete.  Much depends on the states in
the low energy theory, defined by the $n_i$'s.  For example, this theory may itself
have many AdS states, as we will see, which can mean that the system does not reach
the symmetric point.  We will discuss this issue in the models of the subsequent
sections.  Still, we see a sense in which symmetric
points, and especially R symmetric points, are special, and
could plausibly be attractors in a flux landscape.

\section{Model Landscapes}

We are far from understanding the structure of any realistic landscape.  We will
review some of the reasons for this shortly, but just note for now that this
means that any understanding we have of the landscape -- especially
in a cosmological setting -- comes from consideration of model landscapes,
which it is hoped capture some of the important features of what might be
the ``real" situation.

\subsection{The Bousso-Polchinski Model}

The first compelling model of a landscape was that of Bousso and Polchinski (BP)\cite{bp}.
This was based on consideration of four-form fluxes in Type II theories.
One notable
feature of the model
is the assumption that all moduli are fixed for any choice of fluxes, in every state (and fixed,
essentially to the same value).  The energy spectrum is taken to have
the form:
\beq
E_0 = {1 \over 2} N_i^2 q_i^2 - \Lambda_0.
\label{bpenergy}
\eeq
The $q_i$'s are constants, independent of $N_i$.
They are assumed to be small enough that all tunneling amplitudes are small.
This requires (in accord with the estimates in \cite{dfmv})
that the internal manifold be large, with volume scaling as a
positive power of the flux.
This model is extremely useful, first, for illustrating the idea of a
discretuum:  the model
exhibits a nearly continuous distribution of energies for large fluxes.  It also provides
a model for eternal inflation.

With these assumptions, if the $q_i$'s are comparable, transitions tend to decrease
the fluxes uniformly.  One can make a rough estimate of the transition rate by
ignoring gravity and using the formulas appropriate to the thin wall limit\cite{coleman,cdl}.
This was done in \cite{bp}, in the limit that the compactification volume and the radii of
various three cycles are large.  The spirit of our analysis is to consider the case that radii
are of order the fundamental scale, and to examine scaling with $N$.  Then the bubble wall
tensions are of order $1$, while the energy splitting between states is of order $N$.
So the bounce action behaves as $1/N^3$.  As noted in \cite{bp}, even moderately
large volume can significantly suppress the tunneling rate.
Note that
if one of the $N_i$'s becomes much
smaller than the others, further transitions in that flux are suppressed, i.e. all fluxes tend to
decrease uniformly.

\subsection{IIB Compactifications:  Limitations of the BP Model and
The Continuous Flux Approximation}

Known flux models differ from that of BP in some significant ways.  First, when
moduli are fixed, they are typically non-trivial functions of the fluxes (this, and
related points, have been made in \cite{shenker}).
In the IIB theory, for example, the potential for the moduli assumes the usual ${\cal N}= 1$ supergravity form:
\beq
V=e^{K}\left(g^{i\bar{j}}D_iW \bar{D_j W}-3 |W|^2\right)
\label{sugrapot}
\eeq
where $i,j$ are indices labeling the dilaton, complex structure and Kahler moduli.  The superpotential, $W$, is linear in the
fluxes, so
\beq
V = N_i N_j f_{ij}(\phi_\alpha)
\eeq
where $N_i,N_j$ are the fluxes and $\phi_\alpha$ denote the various fields.
If we can think of all fluxes as large, it is natural to write:
\beq
N_i = N u_i.
\eeq
Then, at stationary points of the potential, the $\phi_\alpha$ are independent
of $N$, and the potential is proportional to $N^2$.  But both $\phi_\alpha$, and
$V_0$, the values of the potential at the stationary points, are dependent on the $u_i$'s.
We will illustrate this with some simple toy models shortly.

In the IIB models, there is another important difference:  there are no
flux-independent terms in the classical potential derived from the
supergravity lagrangian.  As a result, there is no obvious identification of the
parameter $\Lambda^2$ or positivity of the parameters
$q_i^2$, and no clear sense in which the fluxes should tend uniformly
towards zero with successive tunneling transitions.

But perhaps the most crucial difference is related to the
Kahler moduli (which we will
denote generally by $\rho$).
The potential for the complex structure
moduli has stationary points, but at large radius of the internal manifold, there
are no stationary points for $\rho$; both perturbatively and non-perturbatively,
the potential tends to zero for large $\rho$.  As a result,
the system is prone to run off to large radius.  The value of $\rho$, and of
the potential (and other quantities) is not similar at the KKLT point and at
adjacent points.
There may be stationary points (metastable vacua of the full quantum system) at small
$\rho$ (i.e. $\rho \sim 1$) for some or most choices of flux (or slightly
larger $\rho$, as in KKLT, see below), but
one would
expect that if one starts in a high cc state with all moduli fixed, one
is likely to tunnel to a configuration which simply rolls out to large $\rho$.
Understanding the significance of this (assuming
it has a sensible interpretation) probably requires a deeper understanding
of eternal inflation, and especially how states are populated.

\subsection{KKLT and Limitations of the Continuous Flux Approximation}

The KKLT model has become a paradigm for the problem of fixing moduli in flux vacua.
While one can raise issues about the analysis (i.e. in what sense the approximations
are truly systematic), it provides a plausible scenario for fixing of moduli in
supersymmetric and approximately supersymmetric IIB compactifications.  It also
provides an illustration of the ways in which realistic landscape states are
unlike the BP model and similar theories which have been used to consider
eternal inflation and related issues.

The basic ingredients in KKLT are a
IIB theory compactified on a CY Manifold (Orientifold).
Before turning on fluxes, the light fields of such a compactification
consist of complex structure moduli and dilaton ($z_i,\tau$),
and Kahler moduli ($\rho_a$).
Fluxes yield a superpotential $W(z,\tau)$.
The equations $D_{z_i} W=0, D_\tau W=0$,
in general, fix $z_i,\tau$.

For large $\rho$,
the low energy effective theory consists of the Kahler
moduli.  In the case of a single
Kahler modulus, KKLT argue
that the superpotential and the
Kahler potential have the form:
\beq W= W_0 + e^{-b \rho};~~K = - 3 \ln(\rho + \rho^\dagger)
\eeq

If there are many states, then, in some, KKLT argued that $W_0$ will be small; in these,
the effective action gives rise to an AdS, supersymmetric minimum with
$\rho \sim -{1 \over b}\ln(W_0)$.
Supersymmetry, they argued,
could be broken by antibranes; alternatively,
dynamics of low energy gauge interactions might provide a breaking mechanism.

The KKLT model makes especially
clear the limitations of
the continuous flux approximation.  The features of KKLT
depend crucially on the smallness of $W_0$, which relies
on cancelations of fluxes.  For nearby choices of flux, $W_0$ is order one
in fundamental units.  So in this instance, we need to step away from the
continuous flux approximation, and consider what the {\it neighboring states}
look like.
These states might then be expected to have broken supersymmetry, with $\rho$ fixed,
if at all, at
some high energy scale, and cosmological constant (positive or negative) also very large.
Moreover, in all of these states, the potential vanishes for very large $\rho$.  So
how transitions might proceed towards the KKLT vacua is not clear.  Transitions from
relatively nearby states, might be expected to end, typically, in big crunches
or with the universe rolling -- eternally -- to large $\rho$.  It could well be that
anthropic considerations permit one to ignore these ``bad ends", but without a larger
theoretical framework, it seems like this will be a challenging story to sort through.

For the $R$-symmetric vacua that we consider shortly, the situation may often be different.
Configurations starting out in a finite neighborhood may well end up
in the preferred state a finite fraction of the time.

\subsection{The Polonyi Model}

Setting aside the problem of $\rho$, if
we consider models which capture more of the features of the IIB theory,
we encounter a richer and more subtle structure.  Consider, first, a familiar
example, the Polonyi model.  We will think of the parameters of
the superpotential of this model as surrogates for flux (we will shortly consider
compactifications with a similar structure, where these parameters are determined
by fluxes).  The model consists of a singlet chiral field $z$, with superpotential:
\beq
W = \alpha + \beta z; ~~K = z z^*.
\eeq
As one varies the parameters, this has both supersymmetric and non-supersymmetric
minima.  In the supersymmetric
minima, $\langle W \rangle$ is not a simple linear function of the fluxes.
$$
z = {-\alpha \pm \sqrt{\alpha^2 -4\beta^2} \over 2
\beta} ;~~ W_0 = {\alpha \pm \sqrt{\alpha^2 - 4 \beta^2} \over 2}
$$

For large $\beta$, there are no supersymmetric solutions.
The non-analytic behavior as a function of the fluxes is not surprising.
If $\alpha = \beta =0$, there is a moduli space of vacua.  Turning on the fluxes
lifts the degeneracy.  This phenomenon is analogous to phenomena in ordinary
degenerate perturbation theory, in which non-analyticity in an expansion
parameter is typical.

The
non-supersymmetric branch exhibits similar behaviors.
We can rewrite the superpotential as
\beq
W = \alpha (1 + u ~z)
\eeq
with $u = \beta/\alpha$; $\alpha$ then plays the role of $n$ in our earlier discussions,
and $u$ of one of the ``angles" introduced there.
To simplify the algebra, we will assume that $\alpha$ and $\beta$ are real; we will also
assume that the vev of $z$, $\langle z \rangle =x$ is real (this preserves a
CP symmetry of the lagrangian).  Then the potential is
\beq
V(x) = \alpha^2 e^{x^2} \left ( u^2 (x^4 - x^2 + 1) + 2u (x^3 -2x) + x^2-3 \right ).
\eeq
The equation, ${\partial V \over \partial x}=0$ then takes the form:
\beq
u^2 (x^5 + x^3) + u (2x^4-x^2 -2) + x^3 -2x = 0.
\eeq
Solving for $x$ in terms of $u$ is difficult, but solving for $u$ in terms of $x$ is
simple.  One root is the supersymmetric branch we discussed above:
\beq
u = -{x \over 1 + x^2}.
\eeq
The second, non-supersymmetric branch has:
\beq
u = -{x^2 -2 \over x^3}.
\eeq

We focus here on the non-supersymmetric branch.  Substituting $u$ into the potential,
yields the value of the energy at the minimum of the potential for a given $x$:
\beq
V_0 = {\alpha^2} {e^{x^2} \over x^6} \left [ x^4 -8x^2 + 4 \right ].
\eeq
Note that the vacuum energy is positive for:
\beq
x^2 > 4 + 2 \sqrt{3}~~~~ x^2 < 4 - 2 \sqrt{3}.
\label{xbounds}
\eeq

We need to check stability; a little bit of algebra yields::
\beq
{d^2 V \over dx^2} \propto {1 \over x^6}[-x^6 + 4x^4 + 12 x^2].
\eeq

Note that this is positive for sufficiently small $x$, and
badly behaved for large $x$.  So we have established the existence
of a range of parameters for which there are stable
stationary points where the vacuum energy is positive and supersymmetry is
broken.


For large and small $u$, we can solve the equations.  In both limits,
\beq
u = -{1 \over x}~~~~~x = -{\alpha \over \beta}
\eeq
For small $\beta$ (large $x$), the vacuum energy is approximately
\beq
V = \beta^2 e^{\alpha^2/\beta^2}.
\eeq
while for small $\alpha$ (small x)
\beq
V = 4 \beta^6/\alpha^4.
\eeq
We see that
the neighborhood of the $\alpha = \beta =0$ point is surrounded by a sea
of negative cosmological constant states with $\alpha$ and $\beta$ comparable.

Thinking of $\alpha$ and $\beta$ as fluxes, we will
take these parameters as large integer
multiples of some basic unit.  We will estimate tunneling rates, as in \cite{dfmv},
by assuming that the tension is proportional to $\Delta x$ in the transition,
and the bounce action is proportional to
\beq
S_b \sim T^4/(\Delta V)^3
\eeq
Then an interesting question to ask is:  starting in a configuration with
$\alpha \ll \beta$, does the system flow towards the region
of AdS states with $\alpha \approx \beta$, or does
it flow towards the zero energy states with $\alpha=0$.
Estimating the bounce action as above, it is easy to see that
\beq
{S_b(\Delta \alpha = 1) \over S_b(\Delta \beta =1)} = {\beta \over \alpha}
\eeq
i.e. the transitions which decrease $\beta$ are faster than those which
decrease $\alpha$.  The couplings tend to equalize, and one is driven to
the negative c.c. regime.

While a simple, toy, model, the Polonyi model already captures certain
features of the IIB flux vacua.  In particular, the expectation values
of moduli fields are complicated functions of the fluxes (both for
supersymmetric and non-supersymmetric vacua).  As a result, the tunneling
behavior of the system is complex, with the system, for example,
tending to avoid the zero c.c. region and heading rather to a region
with negative cosmological constant.  We will encounter behaviors of this
type in a more intricate model in the next section.

\subsection{Multiple-Field Polonyi Model}

In the landscape framework, we are interested in models with {\it many}
fields, analogs of the field $z$.  We can ask whether there are typically
mainly dS or AdS vacua.  A model which is simple to analyze has $P$ fields
and a $S_P$ symmetry:
\beq
W = a\sum_i  z_i + b.
\eeq
To make the analysis simple we will:
\begin{enumerate}
\item  Look for stationary points where the $S_P$
symmetry, and CP, are preserved, i.e. $z_i = x$.
\item  Take the Kahler potential to be
$z_i^* z_i$, and ignore the $e^{K}$ terms in the potential.
\end{enumerate}

We will study the potential in the limit of large $P$.  In this limit, even though the resulting
equations are cubic, we will be able to construct a solution in powers of $1/\sqrt{P}$.
The potential is:
\beq
V =\sum_j (a + z_j^* W) (a + z_j W^*) - 3 W^* W.
\eeq
so
\beq
{\partial V \over \partial z_i^*} = W (a + z_i W^*) +
\sum_j (a + z_j^* W) z_j {\partial W^* \over \partial z_i^*} - 3 W {\partial W \over
\partial z_i^*}.
\eeq
At the would-be minimum,
\beq
W = aPx + b.
\eeq
So, substituting above, we obtain
\beq
0 = P^2 (2 a^2 x^3) + P (3abx^2 - a^2 x) + (b^2 x -2 ab)
\eeq
For large $P$, in order to solve this equation, it is necessary that $x$ take the form:
\beq
x = {\alpha \over P^{1/2}} + {\beta \over P} + \dots.
\eeq
It is easy to see that $\alpha = {1 \over \sqrt{2}}$.
As a result, at the minimum,
\beq
W \approx {a \over \sqrt{2}} P^{1/2}.
\eeq
On the other hand,
\beq
DW \approx a + {1 \over \sqrt{2P}} {a \over {\sqrt 2}} P^{1/2}
\approx {3 \over 2} a
\eeq
so the potential is:
\beq
V = \sum \vert DW \vert^2 -3 \vert W \vert^2 \approx {3 \over 4 } a^2 P.
\eeq
So the solutions are De Sitter, independent of $a$ and $b$.  Including $e^K$, one finds, for large $P$,
that the minima lie near the origin, with cosmological constant
\beq
\Lambda \approx P a^2.
\eeq
Similar results hold for other choices of Kahler potential, such as $K = -\ln(1 + z_i^* z_i)$.

Clearly this is just a model, and one could well imagine that other parameter
choices (and possibly other stationary points) would yield AdS minima, but this
illustrates that there is no particular problem, with large numbers of fields,
obtaining dS spaces.

\subsection{General Orientifolds of IIB on Calabi-Yau Spaces}

In order to relax some of the assumptions of the BP model, it is
useful to consider flux vacua within the context of actual
string compactifications in which the fixing of moduli is more or
less understood.
Perhaps the most widely studied flux vacua are provided by
orientifolds of IIB theories on Calabi-Yau spaces.  Here the candidate
fluxes are RR and NS-NS three form fluxes.  These are constrained
by tadpole cancelation conditions.  Most analyses have
focussed on a low energy supersymmetric effective lagrangian,
obtained by reduction of the ten dimensional supergravity theory.
This lagrangian is described by a Kahler potential
and superpotential for ``light" fields.  There are two classes of
moduli which must be understood:  complex structure moduli, $z_i$, and Kahler moduli,
$\rho_a$.
Classically, the superpotential depends only on the $z_i$ fields.
Typically, one first tries to solve the equations for a supersymmetric stationary point
with respect to the complex structure moduli, $D_{z_i}W=0$.
The Kahler moduli yield a no-scale structure, with vanishing potential.
If the solutions of the equations $D_{z_i}W=0$ yield $W=0$, then the
classical system is fully supersymmetric.  One might expect, for complicated
models, that $W=0$ is the result of an (unbroken) discrete R symmetry.

Our interest, for modeling transitions in the landscape,
lies in non-supersymmetric
solutions.  There are two approaches one might
adopt, closely related to analyses of
Douglas and Denef\cite{douglasdenefnonsusy}.
In one, we will discard the Kahler moduli, keeping only
the complex structure moduli.  In this analysis, non-susy states are stationary
points of the potential, for which $D_i W \ne 0$.  For most choices of flux,
there are, indeed, no supersymmetric solutions.  These models
resemble the Polonyi model, with many AdS minima.  In the second, we keep the
Kahler moduli, using the lowest order Kahler potential, and assuming
that the Kahler moduli are somehow fixed at large values.  SUSY is broken
in all of these states, which are either dS or flat.

\subsection{$T_6$ Inspired Model}
\label{inspired}

Perhaps the simplest example of this type
is the orientifold of the IIB theory on a $T_6$
lattice, studied by \cite{kachrutrivedi}.  This theory is described by a set of complex
structure moduli, $\tau_{ij}$.  Correspondingly, there are a set of NS-NS and R-R
fluxes.  In this section, we will
consider a particular limit of this orientifold, which will allow us to illustrate
certain basic issues surrounding symmetries.
Our principle interest is in non-supersymmetric, dS vacua, which we will
not be able to describe, in general, in any systematic way.  So we will adopt a simple
model.  We will take the radius to be large and fixed, and examine the potential
for the complex structure moduli, ignoring the dilaton.  At the same time, we will include
only the NS-NS fluxes.  In this case the fields of the model are $\tau_{ij}$, $i,j= 1,\dots 3$.
They are described by a superpotential:
\beq
W=a_0 \det(\tau)+a_{ij}({\rm cof}(\tau))_{ij}+b_{ij} \tau_{ij}+b_0
\eeq
and a Kahler potential
\beq
K = \ln(i \det (\tau - \tau^\dagger))
 \eeq

For special choices of the fluxes, the resulting lagrangian exhibits global symmetries.
For example, for
\beq
a_{ij} = a_1 \delta_{ij}~~~~b_{ij} = b_1 \delta_{ij}
\eeq
the lagrangian possesses an $SU(3)$ symmetry.  This symmetry is broken by
the underlying lattice to, at most, some discrete subgroup (we will
discuss the symmetries of this model, in general, shortly), but this breaking is not visible
in the classical theory.  As we will see, such a choice of flux
is enough to fix all of the moduli.

To summarize the model:
\begin{enumerate}
\item  We ignore the dilaton
\item  We keep the Kahler moduli, $\rho_i$, but assume that they are fixed at some large value,
where they can be described by their classical Kahler potential, and $W$ is independent
of $\rho$.  The effect of this is to cancel the $3 \vert W \vert^2$ in the lagrangian
for the $\tau_{ij}$'s.
\item  We work in the continuous flux approximation, ignoring tadpole cancelation
constraints.
\end{enumerate}

Even with these simplifications, the potential for the moduli
is rather complicated to analyze.  In certain limits, however,
it is simple.
An easy case to study is that with $b_0=0$, and $a_1$, $b_1$
proportional to the unit matrix.  With $\tau = i x$, the potential is given by
\beq
V = {3 (a_1^2 x^2 + (b_1 + a_0 x^2)^2 )\over 8x}
\eeq
which has a minimum for
\beq
a_1 = \sqrt{(b_1 - 3 a_0 x^2)(b_1 + a_0 x^2) \over x^2}.
\eeq
At this minimum, one can check that there are no tachyons and the
potential is positive provided
\beq
-b_1 < a_0 x^2 < -{6 b_1 \over 17 + \sqrt{145}}.
\label{abcondition}
\eeq
To understand the significance of this condition, it is useful to express $x^2$ in terms
of $a_1, b_1, a_0$.
\beq
x^2 = {-(a_1^2 + 2 a_0 b_1) \pm \sqrt{(a_1^2 + 2 a_0 b_1)^2 + 12 a_0^2 b_1^2} \over 6 a_0^2}.
\eeq

\subsection{Symmetries of the $T_6$ Model}

Before turning on fluxes, the $T_6$ model has a large moduli space.  On various subspaces,
it exhibits symmetries.  For example, for $\tau$ proportional to the unit matrix,
one has separate $Z_4$ symmetries in each of three planes, as well as an
$S_3$.  In terms of the underlying (real) coordinates, $x^i, y^i,~ i=1,\dots,3$,
the $Z_4$'s take
\beq
x^i \rightarrow y^i~~~~y^i \rightarrow -x^i
\eeq
($i$ fixed).  These are $R$ symmetries, because they change the sign of the holomorphic
three form,
\beq
d\Omega = dz^1 \wedge dz^2 \wedge dz^3
\eeq
with
\beq
z^i = x^i + \tau_{ij} y^j
\eeq
Odd permutations are also $R$ symmetries.

But we need to ask, as well, how the fluxes transform under the various symmetries.
Ref. \cite{kachrutrivedi} provides a convenient catalog of three forms ($H_3(T_6,
Z)$):
\beq
\alpha_0 = dx^1 \wedge dx^2 \wedge dx^3 ~~~~ \beta_0 = dy^1 \wedge dy^2 \wedge dy^3
\eeq
$$~~~~\alpha_{ij} = {1 \over 2}\epsilon_{ik \ell} dx^k \wedge dx^{\ell} \wedge dy^{j}
~\beta^{ij} = -{1 \over 2}\epsilon_{ jk\ell} dy^k \wedge dy^{\ell} \wedge dx^{i}$$
The corresponding fluxes are the quantities
$a_0, b_0, a^{ij},b_{ij}$ encountered above.

It is easy to read off the transformation properties of the fluxes from these
expressions under the various symmetries.  Our particular choice, $a_{ij} = a_1
\delta_{ij}$, breaks all of the $R$ symmetries.  It leaves over cyclic transformations
of the coordinates (i.e. even permutations).

Now consider turning on small off-diagonal entries in $a_1$, $b_1$, breaking the
microscopic cyclic symmetry (in the lagrangian itself, these break the $SU(3)$ symmetry).
For small $a_1$, positive $b_1$ and negative $a_0$,
the condition, eqn. \ref{abcondition},
for the absence of tachyons, reads, approximately:
\beq
-b_1 < -{b_1 \over 3} < -b_1/5
\eeq
which is always satisfied.
Because there are no tachyons, this generates shifts in $\tau_{ij}$ of
order $\delta a_1, \delta b_1$, and changes in the energy of order
$\delta a_1^2, \delta b_1^2$.  Transitions between states are associated
with five branes wrapping three cycles.  If radii are of order one, then the
tensions are of order one (i.e. they are not enhanced by powers of flux).
So, for small $\Delta a_1, \Delta b_1$, transitions towards the symmetry
point are slow; there is no particular tendency of the system to flow towards the
symmetry point; only for very special initial conditions is symmetry achieved.

\subsection{R Symmetries within the $T_6$ Model}

A simple example of an R symmetry in the $T_6$ model is provided by taking a lattice which
is a product of three square lattices in each of three planes.  Consider the symmetry:
\beq
x_1 \rightarrow y_1~~~~~y_1 \rightarrow - x_1
\eeq
This commutes with the orientifold symmetry.  We can classify fluxes according to their behavior under
this symmetry.
The invariant three forms are:
\beq
\alpha_{1 a}~~~\beta^{a1}
\eeq
and
\beq
\alpha_{a1}~~~~\beta^{1a}
\eeq
where $a=2,3$.  The fields $\tau_{1a}$, $\tau_{a1}$ transform like the superpotential,
i.e. by a phase $e^{2 \pi i \over 4}$.
With $\tau_{11} = i + \delta \tau$, $\delta \tau \rightarrow -\delta \tau$
under the symmetry.

To make our expressions simple, we will also impose a symmetry $2 \leftrightarrow 3$
on the fluxes.  Then calling the fluxes (as in
\cite{kachrutrivedi}) $a_{1a}, b_{a1}$, etc
,the superpotential is:
\bea
W= &&a_{12} \left [ \tau_{21} \tau_{33} - \tau_{23} \tau_{31}  + \tau_{31} \tau_{22}
- \tau_{32} \tau_{21} \right ]
+ a_{21} \left [ \tau_{12} \tau_{33} - \tau_{13} \tau_{32}  + \tau_{13} \tau_{22}
- \tau_{23} \tau_{12} \right ]+\cr+&&b^{12}\left[\tau_{12}+\tau_{13}\right]+b^{21}\left[\tau_{21}+\tau_{23}\right]
\eea
We are looking for supersymmetric stationary points with vanishing $W$, so we
wish to solve
\beq
{\partial W \over \partial \tau_{ij}} = 0
\eeq
with $\tau_{11} = \tau_{a1} = \tau_{1a} =0$.  For general choices of the fluxes
$a_{1a}$, etc., the resulting equations, however, are incompatible.
In the language we used in section \ref{rsymmetries}, there are more $X$ type fields ($\tau_{1a},
\tau_{a1}$) than $\phi$ type fields ($\tau_{22},\tau_{33}$).  In more elaborate
models (e.g. Calabi Yau's defined by intersections in weighted projective space)
this is not the case.  In the present model, solutions do exist for special choices
of flux.  In particular, if we take
\beq
b^{12} = b^{21} ~~~a_{12} = a_{21}
\eeq
then there is a moduli space with
\beq
\tau_{22} = \tau_{33}~~~~\tau_{23} = \tau_{32} = \tau_{33} + {b_{12} \over a_{12}}
\label{tauthreetwo}
\eeq

The next step is to turn on small symmetry violating fluxes, fixing these remaining
moduli, and studying the behavior of the energy and fields as functions of the small fluxes.
To first approximation, we simply need to write the effective action for the light
moduli, and find its stationary points.
There are many fluxes which can be turned on, and the resulting potentials, for which
supergravity corrections are important, are complicated.
To get some flavor for the problem,
note that we need to fix $\tau_{11},\tau_{22}$ and $\tau_{33}$.  So we can turn on
small
fluxes $a_0$, and similarly $b_{11},b_{22},b_{33}$ (again, to keep things
simple, we will impose a $1 \leftrightarrow 2$ symmetry).  Then the added terms in the
superpotential are:
\beq
W_{l.e.}= a_0 (\tau_{11} \tau_{22} \tau_{33} -
\tau_{11} \tau_{23} \tau_{32}+ \dots) + b_{11} \tau_{11} + b_{22}(\tau_{22}
+ \tau_{33}).
\eeq
One can check that, with these choices, there are no tadpoles for $\tau_{1a}$, $a=1,2$.
We can substitute for $\tau_{23},\tau_{32}$ their forms on the moduli space,
eqn. \ref{tauthreetwo}.  The resulting expression is similar to that we studied in
section \ref{inspired}, and we expect a similar structure for the solutions.
Energies are of order $n^2$ (where the $n$'s are the symmetry-breaking fluxes above).
In particular, we expect that the diagonal elements of $\tau$ are fixed,
that for a range of fluxes there are no tachyons, and that transitions tend to
keep the diagonal elements of $a,b$ comparable.  So shifts of the heavy fields will
be of order $n^2/N^2$.


Now we see that the tunneling rates are as we guessed for $R$-symmetric stations
in section \ref{attractors}.  Energies have the form:
\beq
E_0 = a n^2 + b {n^4 \over N^2}
\eeq
so energy splittings between states of different $N$ behave as $n^4/N^3$.
The tensions, again, are of order one, parametrically, the bounce actions for transitions which change $N$ behave as
\beq
S_b(\Delta N=1) \approx {N^9 \over n^{12}}
\eeq
On the other hand, transitions which change $n$ simply scale as
\beq
S_b(\delta n=1) \approx {1 \over n^3}.
\eeq
So symmetry-restoring transitions are highly favored.

\section{Conclusions}

Even if underlying fundamental physics is some type of landscape of metastable vacua,
it is unlikely that, in the near future, we will have anything resembling a complete catalog
of these states.  Instead, any sort of understanding of the connection of microphysics and
the phenomenon we observe will have to be based on robust statements about distinctive
features of physics at much lower energy scales.   In an earlier paper\cite{dfmv},
we focussed on supersymmetry and the question of
stability.  In the present paper, we have
mainly considered discrete symmetries.  We have remarked that states with
discrete symmetries are rare\cite{dinesun}, but suggested that they might be cosmological
attractors.  This was motivated by the naive assumption that the potential might
grow quadratically with flux around the symmetric point, so the system, starting in the
neighborhood of the symmetric point, might tend to tunnel towards the symmetric point.
To study this problem we have considered several {\it model landscapes}, which capture
likely features of a would-be underlying cosmic landscape.  Even before approaching
the problem of symmetries, we have argued that the Bousso-Polchinski model (the simplest
model landscape we have considered) is likely too simple a picture, given
its assumptions
about positivity and constancy of flux coefficients,
and given the existence of asymptotic pseudomoduli
spaces.  We have considered a number of toy models which capture more, but not all, of
these expected features.

To consider the question of whether symmetric points are attractors, we have worked
in the continuous flux approximation.  This is hardly likely to be realistic, but it should
give some feeling of whether it is common for the system to reach symmetric points.
Again, it provides a model, relatively simple to analyze, in which one can assess whether
there is or is not a net drift towards the symmetric points.
We have seen that plausibly an order one fraction of states lie within the neighborhood
of symmetric points.  However, we have also seen that for non-R symmetries, this fact is
of no particular importance.  Even if the system starts in the neighborhood, the system will
drift towards non-symmetric, AdS states.

For R symmetric states, the situation is potentially quite different.  We have investigated,
first, the question of whether there are likely to be predominantly dS or AdS states in the
neighborhood.  Within the models we have studied, both situations arise, so it seems
plausible that, for a finite fraction of R symmetric points, the neighborhood is populated
by metastable dS states.  In this case, we have seen, that typical states in the neighborhood
{\it will} drift towards a symmetric state.  We take this to mean that an order one fraction
of possible starting points leave the system in a symmetric state.  Thus, even though
such states are rare, cosmological considerations are likely to favor them.

Based on this study, we offer the following tentative conclusions:
\begin{enumerate}
\item  An underlying landscape is likely to lead to low energy supersymmetry, simply
because supersymmetry seems the most generic way to guarantee stability.
\item  Discrete R symmetries might plausibly emerge in a landscape.
Even though these states are far less numerous than states without such symmetries, they
are likely
to be cosmological attractors in a finite fraction of cases.  As discussed in \cite{branches},
they should tend to exhibit supersymmetry breaking at low energies, and there can easily
be enough of them to resolve the cosmological constant problem.
\item  Discrete non-R symmetries (as one might want to understand, say, the structure of the
NMSSM or some flavor structure) do not seem favored in any particular way.  They are rare,
and don't seem to be singled out by (at least naive) cosmological considerations.
\item  Slightly increasing the size of the discrete symmetry does not seem to lead
to a significant suppression, so intricate discrete symmetries (such as might suppress
dimension five proton decay) might plausibly arise.
\end{enumerate}

As we have explained, establishing these statements as an ironclad ``top down" prediction
is difficult, but it seems likely that further landscape studies could well establish that
they are robust.

In this paper, we have focussed principally on discrete symmetries.  The problem of gauge
symmetries is of obvious interest.  In field theory, it is well-known that finite temperature
tends to favor the maximal unbroken gauge symmetry.  The issues, in the landscape context,
are different and somewhat more complicated than those connected with discrete symmetry,
and require a separate investigation, which will be reported elsewhere.

\clearpage

{\bf  Acknowledgements}

We acknowledge valuable conversations with Tom Banks, Ben Freivogel, and Eva Silverstein.  We thank
Matt Johnson for sharing with us some of his concerns about IIB theories and
eternal inflation, which
bear some similarity to issues raised here, and Leonard Susskind
and Steve Shenker for asking us several important, focussed
questions.  This work supported in part by the U.S.
Department of Energy.

\end{document}